\documentclass[twocolumn, amsmath,amssymb,superscriptaddress,prb]{revtex4}
\usepackage{graphicx}
\usepackage{bm}
\usepackage{subfigure}
\usepackage{multirow}
\usepackage{color}

\begin{document}


\title{Dirac cone shift of a passivated topological {Bi$_2$Se$_3$} interface state}

%



\author{Gregory S. Jenkins}
    \homepage{http://www.irhall.umd.edu}
    \email{GregJenkins@MyFastMail.com}
    \affiliation{Department of Physics, University of Maryland at College park, College Park, Maryland, 20742, USA}
    \affiliation{Center for Nanophysics and Advanced Materials, University of Maryland at College park, College Park, Maryland, 20742, USA}
\author{Andrei B. Sushkov}
    \affiliation{Department of Physics, University of Maryland at College park, College Park, Maryland, 20742, USA}
    \affiliation{Center for Nanophysics and Advanced Materials, University of Maryland at College park, College Park, Maryland, 20742, USA}
    \affiliation{Materials Research Science and Engineering Center, University of Maryland at College park, College Park, Maryland, 20742, USA}
\author{Don C. Schmadel}
    \affiliation{Department of Physics, University of Maryland at College park, College Park, Maryland, 20742, USA}
    \affiliation{Center for Nanophysics and Advanced Materials, University of Maryland at College park, College Park, Maryland, 20742, USA}
\author{Max Bichler}
    \affiliation{Walter Schottky Institut and Physik Department, Technische Universitat Munchen, D-85748 Garching, Germany}

\author{Gregor Koblmueller}
    \affiliation{Walter Schottky Institut and Physik Department, Technische Universitat Munchen, D-85748 Garching, Germany}
\author{Matthew Brahlek}
    \affiliation{Department of Physics and Astronomy, The State University of New Jersey, Piscataway, New Jersey 08854, USA}
\author{Namrata Bansal}
    \affiliation{Department of Physics and Astronomy, The State University of New Jersey, Piscataway, New Jersey 08854, USA}
\author{Seongshik Oh}
    \affiliation{Department of Physics and Astronomy, The State University of New Jersey, Piscataway, New Jersey 08854, USA}
\author{H. Dennis Drew}
    \affiliation{Department of Physics, University of Maryland at College park, College Park, Maryland, 20742, USA}
    \affiliation{Center for Nanophysics and Advanced Materials, University of Maryland at College park, College Park, Maryland, 20742, USA}
    \affiliation{Materials Research Science and Engineering Center, University of Maryland at College park, College Park, Maryland, 20742, USA}
\date{\today}

\begin{abstract}
Gated terahertz cyclotron resonance measurements on epitaxial Bi$_2$Se$_3$ thin films capped with In$_2$Se$_3$ enable the first spectroscopic characterization of a single topological interface state from the vicinity of the Dirac point to above the conduction band edge. A precipitous drop in the scattering rate with Fermi energy is observed that is interpreted as the surface state decoupling from bulk states and evidence of a shift of the Dirac point towards mid-gap. Near the Dirac point, potential fluctuations of 50 meV are deduced from an observed loss of differential optical spectral weight near the Dirac point. Potential fluctuations are reduced by a factor of two at higher surface Fermi levels in the vicinity of the conduction band edge inferred from the width of the scattering rate step. The passivated topological interface state attains a high mobility of $3500$ cm$^2$/V$\cdot$s near the Dirac point.
\end{abstract}

\maketitle

The topological insulator (TI) is a unique new state of matter that is a bulk insulator possessing a topologically protected metallic surface state of massless particles known as Dirac fermions. Key properties of this surface state are spin-momentum locking, suppression of back scattering, and an intrinsic magneto-electric effect.\cite{HasanKaneRMP2010,QiZhangRMP2011}  The experimental realization of this state of matter has sparked great interest owing not only to their potential use in spintronics and quantum computing but in the investigation of the fundamental nature of topologically nontrivial quantum matter.\cite{DrewZhangPRL2010, Qi_Zhang_Kerr_2008}

The surface state Dirac point of most currently known TIs lies below the top of the bulk valence band (like Bi$_2$Te$_3$).\cite{HasanKaneRMP2010,QiZhangRMP2011} This deficiency is particularly problematic since many interesting phenomena are predicted to occur at or near the Dirac point, like the 1/2-quantized Hall step.\cite{HasanKaneRMP2010,QiZhangRMP2011} In this regard, Bi$_2$Se$_3$ is one of the few attractive materials since its single Dirac point lies well above the valence band.\cite{XiaARPESNP2009} However, Bi$_2$Se$_3$ is especially prone to defect doping in the bulk which dominates the conduction.\cite{ButchPRB2010, Jenkins_PRB2010,  Analytis_NP2010,SteinbergHerreroNNano2010} Surface doping from defects and atmospheric exposure not only lower the mobility but result in accumulated surfaces that inhibit conventional gating techniques from decoupling the surface states from bulk states.\cite{Oh_arxiv2011, KongARPESNN2011,SteinbergHerreroNNano2010,Checkelsky_OngPRL2011}

\begin{figure}
\includegraphics[scale=.40]{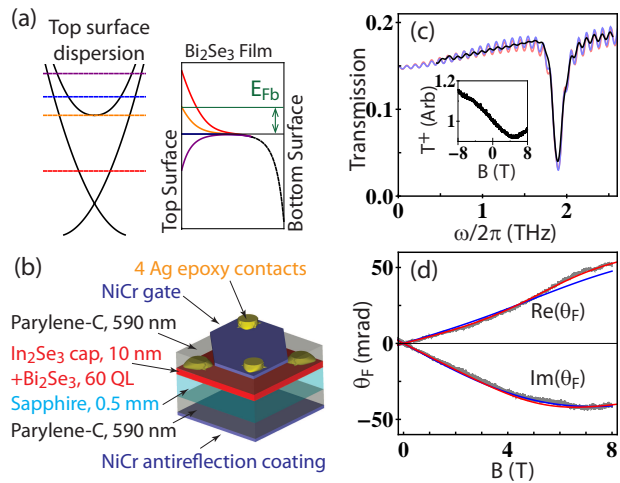}
\caption{\label{fig1}\textbf{Zero-gate normal incident transmission measurements} (a) A band bending schematic showing the conduction band edge (CBE) relative to the bulk chemical potential (green), and the corresponding Dirac cone Fermi level associated with the vacuum interface, when a variable gate voltage is applied to the top surface.  Purple through red correspond to increasingly negative gate voltages. The bottom surface (black) is accumulated due to defects. (b) A schematic represents the measured device. (c) Measured 6 K Fourier transform (FTIR) transmission spectrum (black). A single-fluid Drude model fit (blue) gives $\gamma=1/\tau=7.3$ THz and  $\omega_{ps}/(2 \pi)=27.9$ THz that translates to a carrier density $n=1.23\times 10^{13}$ cm$^{-2}$ for a mass of $m=0.21$. Oscillations in the FTIR data are due to Fabry-Perot etalon in the sapphire substrate. (inset) Measured 10 K cyclotron resonance transmission at $\omega/(2 \pi)$=0.74 THz. (d) Complex Faraday angle measured at 10 K and $\omega/(2 \pi)$=0.74 THz (black). A single-fluid Drude model fit (blue) gives $\gamma=1/\tau=7.4$ THz, $n=1.1\times 10^{13}$ cm$^{-2}$, and $m_c=0.21$.  The red curves in figures (c) and (d) are computed from a three-fluid Drude model with parameters reported in Figure \ref{fig3}.
}
\end{figure}

Passivating and controlling the electronic properties of the TI surface state involves interfacing with other materials. The proximity of the topological insulator to trivial insulators, magnetic materials, and superconductors are either expected to induce changes in the Dirac cone\cite{KorenmanDrew1987,AgassiKorenman1988} or fundamentally change the nature of the interface state giving rise to exciting new emergent phenomena.\cite{HasanKaneRMP2010,QiZhangRMP2011,DrewZhangPRL2010, Qi_Zhang_Kerr_2008} The usual surface sensitive techniques that have successfully characterized the vacuum interface are not useful when the topological interface state is deeply buried beneath other materials.\cite{HasanKaneRMP2010,QiZhangRMP2011,RoushanSTMYazdani_2009, BeidenkopfYazdaniNP2011} The interpretation of dc transport measurements of the TI surface state is confounded by the multiple conductivity channels generally expected in TIs: a bulk contribution with two inequivalent surfaces, each with a topological surface state and bulk screening surface layer (depicted in Figure \ref{fig1}(a)). To characterize these systems, new experimental techniques are required.

In this letter, all of the above issues are experimentally addressed. A thin film of Bi$_2$Se$_3$ is epitaxially grown on sapphire and capped with In$_{2}$Se$_{3}$.\cite{Bansal_OhThinFilm2011} The $\alpha$-phase crystalline In$_{2}$Se$_{3}$ cap\cite{OhBiInSePRL2012} protects the Bi$_2$Se$_3$ from atmospheric degradation and dopes the surface enabling the Dirac point to be reached by conventional gating while achieving high mobilities.  Evidence of a large shift of the Dirac point towards the conduction band edge relative to the vacuum interface, due to the In$_{2}$Se$_{3}$ capping layer, is reported for the first time demonstrating the possibility of controlling the Dirac cone in other TI systems. The first experimental determination of the surface state scattering rate as a function of Fermi energy provides a detailed map of the interplay between the bulk and TI surface state carriers. The energy scale of potential fluctuations characterized near the Dirac point agrees with other measurement techniques,\cite{Dohun_Fuhrer2012,BeidenkopfYazdaniNP2011} and is reduced by a factor of two near the conduction band edge.

Using terahertz cyclotron resonance measurements, each conduction channel in the film is a distinct Lorentzian response distinguishable by the sign of the charge, cyclotron mass $m_c$, spectral weight $n e^2/m_c$, and carrier scattering rate $\gamma$. Spatial location of the carrier contributions in the film are ascertained by concurrently modulating a semi-transparent top gate that spatially modulates the charge distribution in the film in a predictable way.

Zero-gate normally incident transmission measurements are shown in Figure \ref{fig1}(c,d) for the device schematically represented in Figure \ref{fig1}(b). The Fourier transform spectroscopic transmission measurement of Figure \ref{fig1}(c) was performed in zero magnetic field with unpolarized incident light. The zero-gate cyclotron resonance transmission measurement shown in the inset of Figure \ref{fig1}(c) was performed with circularly polarized incident light at a fixed laser frequency as a function of magnetic field applied normal to the film. A Lorentzian absorption is expected when the probe frequency equals the cyclotron frequency $\omega_c = e B/m_c$.  The resonant position located in positive magnetic field corresponds to n-type carriers of mass $m_c\approx0.19$ $m_0$. The complex Faraday angle $\theta_F$ reported in Figure \ref{fig1}(d) was measured with polarization-modulated incident light at a fixed frequency, a method detailed elsewhere.\cite{Jenkins_RSI_2010} Fits to the zero-gate transmission data in Figure \ref{fig1}(c,d) using a single-fluid Drude model for the conductivity, depicted by the blue curves, reveal a large carrier density $n\gtrsim10^{13}$ cm$^{-2}$ and a dominant cyclotron mass that is much larger than the conduction band edge mass $m_c=0.15$ $m_0$ measured in bulk single crystals.\cite{Sushkov_PRB2010,Jenkins_PRB2010,ButchPRB2010,AguilarPRL2012, Köhler1973, SM} Deviations between the $\theta_F$ data and fit are outside of measurement error, hinting at small contributions from other conducting channels.

\begin{figure*}
\includegraphics[scale=.75]{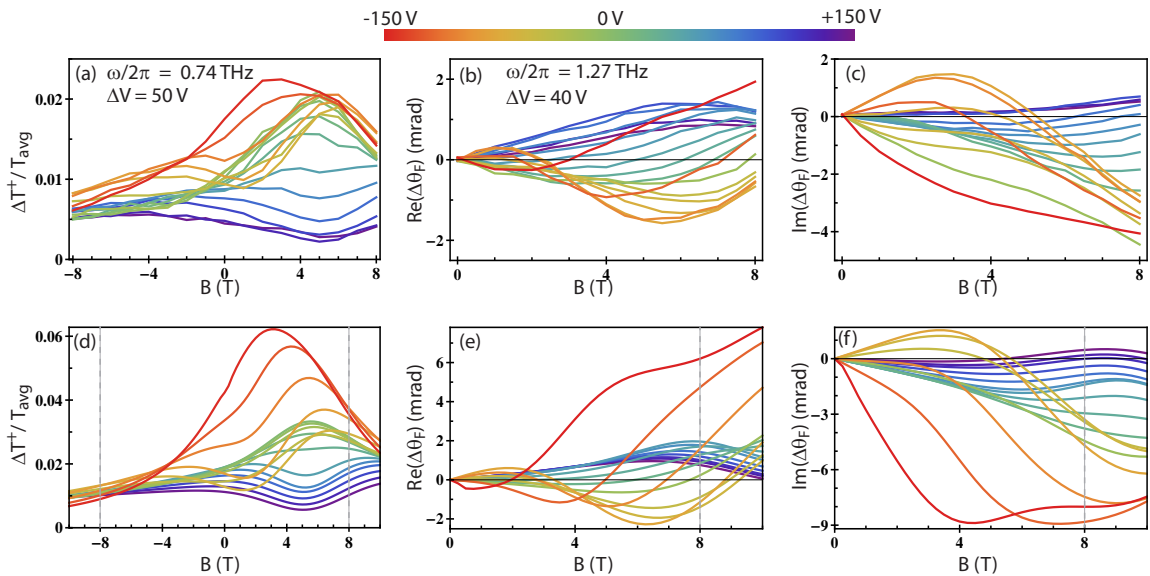}
\caption{\label{fig2}\textbf{Gated measurements at 10K:} All optical signals are differences between two gate values, $V_g \pm \Delta V/2$. Average gate values $V_g$ for each plot are coded according to the colored bar at the top of the figure.  Shown on each data graph is the radiation frequency $\omega/(2 \pi)$ and $\Delta V$ value. (a) Differential cyclotron resonance data normalized to the average with a standard deviation of the mean of $5\times 10^{-4}$  measured at every integer B-field. (b,c) Differential Faraday angle data measured at every 1/2-integer B-field with a standard deviation of the mean of $30$ $\mu$rad (d-f) The color of each curve of the modeled optical response correspond to the same $V_g$ and $\Delta V$ values appearing in the respective data sets (a-c).
}
\end{figure*}

Modulating the top gate voltage at $\sim 1$ Hz and measuring the difference in cyclotron resonance ($\Delta$-CR) reveals the top topological interface state properties. The largest negative average gate voltage in Figure \ref{fig2}(a) shows a resonant component at $B\approx2$ T, corresponding to a cyclotron mass much less than the bulk mass.\cite{Sushkov_PRB2010,Jenkins_PRB2010,ButchPRB2010} Since the cyclotron mass $m_c = \hbar k_F/v_F$ is expected to go to zero as the Dirac point is approached, such a small measured cyclotron mass is direct evidence of the topological interface state. The corresponding  Fermi level is approximately 50 meV above the Dirac point, far below the CBE of 190 meV observed on Bi$_2$Se$_3$.\cite{ZhuPRL2011,ZhuPrivate} Even though a more sophisticated analysis of the scattering rate will be given later, an estimate for the feature at $2$ T  taken directly from the raw data is given by the full width at half maximum, $\Delta B/B_{res}=2\gamma/\omega \approx  4$, so the inverse lifetime $\gamma=1/\tau$ is in the vicinity of $9$ THz.

As depicted in Figure \ref{fig1}(a) for a surface Fermi level near the Dirac point (red), a depletion region necessarily exists with a depletion charge density $e n_d$. As the gate increases, the surface state and depletion region begin filling (orange), eventually leading to an accumulated layer (purple). From the Dirac point to $V_g=0$, $4.4\times10^{12}$ cm$^{-2}$ carriers are added to the film.  The net carrier density of the top surface is $n_s=4.4\times10^{12} \text{ cm}^{-2} - n_d < 4.4\times10^{12} \text{ cm}^{-2}$. By estimating $n_d$, $n_s$ is $<3\times10^{12}$ cm$^{-2}$.\cite{SM} The large majority of the carriers measured at zero-gate therefore reside in the bottom half of the film.

Transport measurements on similarly grown films, performed over a wide range of thicknesses, characterize the bulk conductivity which is small compared to surface contributions.\cite{Oh_arxiv2011} This measured bulk density is a small fraction of the zero-gate optically measured carrier density. Therefore, a significant portion of the zero-gate carrier density likely resides in a bottom surface accumulation layer that includes a topological surface state. However, two distinct large spectral weight contributions with different masses are not discernible in the zero-gate broad responses reported in Figure \ref{fig1}. The differential measurements are insensitive to carrier contributions that are not significantly modulated by the top gate and are therefore expected to be very insensitive to the bottom surface properties.

The $\Delta$-CR measurements are sensitive to the absorptive (real) part of the conductivity. Additional information is attained from the differential complex Faraday angle data reported in Figures \ref{fig2}(b,c). The real (imaginary) part of $\theta_F$ is related to the reactive (absorptive) part of the conductivity.\cite{Jenkins_RSI_2010}

To gain further insight into the gate-dependent data, a model is developed having only one assumption: the top TI interface state is presumed to be described by the Dirac cone characterized by angular resolved photo-emission (ARPES).\cite{ZhuPRL2011,ZhuPrivate} Deviations between the model and data therefore provide critical information about how the spectroscopically measured interface state properties differ from the ARPES measured Dirac cone. There are two components to the model: Thomas-Fermi screening, and optical transmission with the conductivity expressed in Drude form. Only three free parameters enter into the differential optical models.

A Thomas-Fermi (T-F) screening model is used to reduce the number of free parameters, which requires only the value of the top surface Fermi level produced by a specified gate voltage and the bulk carrier density. With this information, the amount of charge a gate moves into the top TI surface state and the rest of film are both determined. The bulk density is set to $5\times10^{17}$ cm$^{-3}$ consistent with dc transport on similarly grown films.\cite{Oh_arxiv2011}  The optical model results are not very sensitive to changes in this parameter. The small measured mass determines the surface Fermi level for a specific gate voltage.

Three Drude conductivity contributions are incorporated into the optical transmission formulas. Only two contributions are gate dependent, the top topological surface state (TSS) and the modulated top bulk (MTB) region near the surface. As depicted in Figure \ref{fig1}(a), the MTB region includes a gate-dependent depletion region and screening region (red) for surface Fermi levels below the conduction band edge (orange), which gradually become a constant bulk contribution (blue) plus a changing accumulation region (purple) at higher surface Fermi levels. Each Drude contribution has three parameters: $n$, $\gamma$, and $m_c$. For the TSS, the only free parameter is $\gamma$ since $n$ is determined from the T-F model and $m_c$ is determined from the ARPES Dirac cone dispersion. There are two MTB free parameters, $m_c$ and $\gamma$, whereas $n$ is determined from the T-F model and defined as zero at the Dirac point.

The remaining Drude term is gate-independent and the largest conductivity contribution in the film, which is conceptualized as a bottom surface accumulation layer (BSAL).\cite{SM}

Figures \ref{fig2}(d-f) show the modeled differential optical signals, and Figures \ref{fig1}(c,d) (red curves) show the modeled zero-gate signals. Comparisons with other $\Delta$-CR  measurements at higher frequencies as well as gated FTIR spectroscopy data are shown in the Supplemental Materials.\cite{SM} All the features are reproduced over a wide range of applied voltages, magnetic fields, and frequencies, but interesting deviations appear at high negative gate voltages where the top TI surface state carrier density is small.

\begin{figure}
\includegraphics[scale=.4]{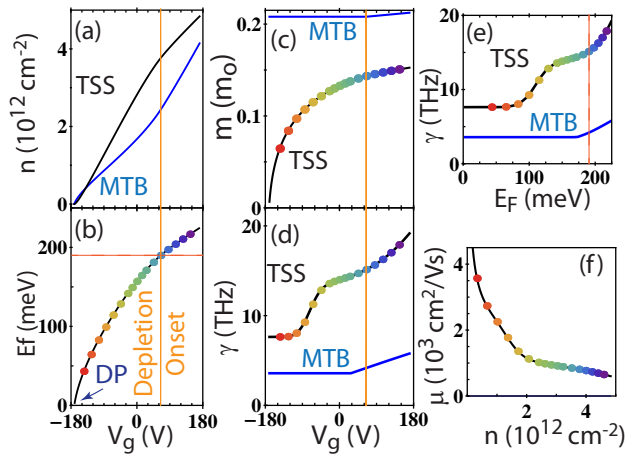}
\caption{\label{fig3}\textbf{Model parameters:} Three n-type Drude terms model the reported differential optical measurements. The BSAL parameters are $n=1.0 \times 10^{13} $ cm$^{-2}$, $ m_c=0.25$ $m_0$, and $\gamma=1/\tau=9.4$ THz.  The red through purple dots on all graphs depict the same average voltages applied as in Figure \ref{fig2}(a). The solid orange lines depict the depletion onset and the red dashed lines depict the conduction band edge (CBE), inferred from the ARPES measured dispersion.(a) The fraction of charge moved by the gate in the MTB (blue) carrier contribution and the TSS (black), where the flat-band bulk density is $5\times10^{17}$ cm$^{-3}$, shown as a function of $V_g$. (b) The TSS Fermi level with respect to the Dirac point. (c) The cyclotron mass in units of the bare electron mass $m_0$ and (d) the scattering rates (inverse lifetimes) are shown for the MTB (blue) and TSS (black) carrier contributions. (e) Scattering rates shown as a function of the TSS Fermi level. (f) The mobility, calculated from the mass and scattering rate versus the TSS number density. }
\end{figure}

The BSAL Drude parameters are determined from zero-gate data fitting. For surface Fermi levels below the CBE,
a constant $\gamma$ and $m_c$ for the MTB carriers best fit the differential data. For this especially important range, the only remaining variable gate-dependent free parameter is $\gamma$ of the TI interface state. Figure \ref{fig3} summarizes the Drude parameters which most closely reproduce the data sets.

The peak-dip-hump structure of the $\Delta$-CR model in Figure \ref{fig2}(d) at negative gate voltages are caused by the rapidly increasing TI interface state scattering rate with surface Fermi energy, while the shifting structure to higher B-field ($m_c$) is consistent with the ARPES measured Dirac cone spectrum. The model is extremely sensitive to the functional form of the TSS scattering rate allowing accurate determination of this parameter. A thorough intuitive understanding of the $\Delta$-CR model output in relation to Drude parameters are presented in the Supplemental Materials.\cite{SM}

However, deviations in magnitude between the $\Delta$-CR model and data for $V_g \lesssim -100$ V become progressively larger as the Dirac point is approached, indicating the presumed ARPES Dirac cone overestimates the $\Delta$-CR spectral weight of the TI interface state. The $\Delta$-CR signals are proportional to changes in the optical spectral weight, $n e^2/m_c$. The ideal Dirac cone dispersion predicts $n/m_c \propto k_F v_F$ which goes to zero at the Dirac point as do both $n$ and $m_c$.

However, potential fluctuations are not included in the Dirac cone model. In the presence of potential fluctuations,\cite{DasSarmaGraphenReview2011,Dohun_Fuhrer2012,BeidenkopfYazdaniNP2011,ChenFuhrerNPhys2008} the spectral weight saturates to some non-zero rms value as the Dirac point is approached. This saturation of the spectral weight reduces the observed differential spectral weight below that predicted by a Dirac cone dispersion.

Within this interpretation, these deviations provide a means to estimate the characteristic potential fluctuation energy. The first significant deviation between model and data of Figure \ref{fig2}(a,d) begins with average voltage $V_g=-105$ V with its associated lowest voltage of $-130$ V. From Figure \ref{fig3}(b), the TSS Fermi level corresponds to $\sim 60$ meV above the Dirac point. This estimate of the potential fluctuations agrees well with other estimates.\cite{Dohun_Fuhrer2012}

The TSS scattering rate, shown in Figure \ref{fig3}(e), monotonically increases with surface Fermi level. The mobility is calculated from the mass and scattering rate, reported in Figure \ref{fig3}(f).

The scattering rate shows a step at $-70$ V, well below the inferred gate voltage of the CBE that is 190 meV above the Dirac point as measured by ARPES. This behavior is very different from that observed in graphene.\cite{HorngGrapheneGammaPRB2011,DasSarmaGraphenReview2011} The step in scattering rate may relate to the decoupling of the topological surface state from bulk scattering channels. This is expected to occur when the surface and bulk states become non-degenerate at surface Fermi levels below the CBE when a depletion layer begins forming between the bulk and surface carriers. In this interpretation, the step in scattering rate is the optical signature of the CBE. No such signature is present near 190 meV.  For a shifted Dirac cone where the CBE is at $V_g=-70$ V, the corresponding surface Fermi level is calculated from Thomas-Fermi screening arguments, similar to the results of Figure \ref{fig3}(b). The CBE is $\sim80$ meV above the Dirac point (and a refined estimate of the potential fluctuation onset at $V_g=-130$ V is $\sim$50 meV).\cite{SM} The interface state Dirac point appears shifted by $\sim 110$ meV relative to the vacuum interface.

Such a shift of the Dirac point is not unexpected for the In$_2$Se$_3$/Bi$_2$Se$_3$ interface.\cite{KorenmanDrew1987, AgassiKorenman1988}  The shift depends upon the size of the band gap as well as the relative alignment of the bands that in part depend upon the electron affinities and any potential steps caused by interface states.\cite{Ruan_Ching_1987}  These parameters are not well known for In$_2$Se$_3$ or Bi$_2$Se$_3$. However, an upward shift of $\sim 110$ meV in the position of the Dirac point for the In$_2$Se$_3$/Bi$_2$Se$_3$ interface is reasonable.\cite{SM} Within this picture, the location of the step in scattering rate is naturally explained, and the width of the step is a consequence of potential fluctuations, $\sim 25$ meV. Enhanced screening from a higher density of surface carriers as well as contributions from bulk carriers tend to reduce the potential fluctuations compared to the Dirac point vicinity.\cite{DasSarmaGraphenReview2011,CulcerStanescuDSarmaPRB2010}

The gate modulated spectroscopic techniques presented here provide a powerful means to characterize the topological surface state even when beneath other materials and in the presence of multiple conduction channels. The possibility of passivating the surface and shifting the Dirac point by capping provides a new flexibility in tailoring the topological surface states.


The authors thank M. S. Fuhrer, Dohun Kim, T. D. Stanescu, and S. Das Sarma for helpful conversations, and A. Damascelli and Z.-H. Zhu for providing the ARPES data for the TSS dispersion of Bi$_2$Se$_3$. The UMD work is supported by DOE DE-SC0005436 and CNAM. The Rutgers work is supported by IAMDN, NSF DMR-0845464, and ONR N000140910749.


\bibliography{BiSefilmBib}


\newpage
\setcounter{figure}{0}
\newpage
\section*{SUPPLEMENTAL MATERIAL\label{sec:sup:title}}

Experimental methods and models presented in the main text are described in more detail. More supporting data with comparisons to the same model are reported. Important issues relating to statements in the main text are discussed in greater depth.

The sample and device characteristics as well as miscellaneous experimental details are discussed in Section \ref{sec:sup:methods}. The measured ARPES surface dispersion and a summary of the Thomas-Fermi screening and optical models are presented in Section \ref{sec:sup:Models}. In Section \ref{sec:sup:Discussion}, discussions are presented regarding an intuitive understanding of the $\Delta$-CR model of Figure 2(d), examination of carrier distributions in the film and energy estimates of the conduction band edge and potential fluctuations, the discrepancy between the non-parabolicity implied by the mass as measured by Shubnikov-de Haas in comparison to cyclotron resonance, and the shift of the Dirac cone in the Bi$_2$Se$_3$/In$_2$Se$_3$ interface state relative to the Bi$_2$Se$_3$/vacuum state. Additional gated data, including $\Delta$-CR, FTIR spectroscopy, and $\Delta\theta_F$ data, is presented in Section \ref{sec:sup:moredata}.

\section{Methods\label{sec:sup:methods}}

A Bi$_2$Se$_3$ 60 quintuple layer film was grown epitaxially onto a 0.5 mm thick sapphire substrate of area 1x1 cm square.\cite{Bansal_OhThinFilm2011} A 10 nm thick In$_{2}$Se$_{3}$ capping layer was immediately deposited without breaking vacuum.

The In$_{2}$Se$_{3}$ layer is verified to be crystalline $\alpha$-phase via RHEED measurements as described in reference \citenum{OhBiInSePRL2012}. In this article, the RHEED image in Figure 1(a) shows that  (Bi$_{1-x}$In$_x$)$_2$Se$_3$ for $x=0.05$ is a single phase. When the film is very thin (10 QL or so), distinguishing between $\alpha$ or $\gamma$ phases, or some other crystalline phase, is difficult. But even for such thin films, RHEED shows the existence of only a single phase. The $\alpha$-phase is closest to Bi$_2$Se$_3$ in structure, and it is well known that even 1\% of Bi in In$_{2}$Se$_{3}$ stabilizes the $\alpha$-phase over other competing phases. Although the data is not shown in reference \citenum{OhBiInSePRL2012}, we find that the phase as measured by RHEED on 100\% In$_{2}$Se$_{3}$ on Bi$_2$Se$_3$ looks identical to $x=0.05$ (Bi$_{1-x}$In$_x$)$_2$Se$_3$.  It appears that the base layer of Bi$_2$Se$_3$ provides a stabilizing seeding template for the In$_{2}$Se$_{3}$ $\alpha$-phase.

Two contacts were made to the film using Eccobond Solder 59C. Parylene-C was deposited to a thickness of $590$ nm, conformally encasing the sample on both the top and bottom surfaces with a measured thickness uniformity better than $10$ nm. NiCr films were evaporated onto the parylene-C, the bottom surface serving as an absorptive broadband antireflection coating and the top surface as a gate patterned by shadow masking.  The antireflection coating was deposited to a sheet resistance of $275$ $\Omega/\square$ and the gate to $400$  $\Omega/\square$. Two contacts to the gate were made using the same silver epoxy.

Parylene-C optical properties were characterized utilizing FTIR transmission measurements. The static dielectric constant was measured with a capacitance bridge and found to be 2.8 for parylene-C at low temperature consistent with reference \citenum{Kahouli_paryleneC_2009}. The gate provides $2.6 \times 10^{10}$ e/cm$^2$  per volt. Breakdown voltages as large as 200 V were achieved.

FTIR measurements were performed with a Bomem DA8.  A far-infrared laser cavity pumped by a CO$_2$ laser provided the fixed frequency source for the Faraday and cyclotron resonance measurements. The polarization modulation technique used to measure the Faraday angle is detailed elsewhere.\cite{Jenkins_RSI_2010} Circular polarized light for the cyclotron resonance measurement was generated using quartz quarter-waveplates with NiCr antireflection coatings.  A GaAs 2-DEG heterostructure was used to set the waveplate angle, verify the retardance of the waveplates, and served to calibrate the Faraday angle measurements and the sign of the charge carriers in the cyclotron resonance measurements.




\renewcommand{\thefigure}{S\arabic{figure}}
\setcounter{figure}{0}

\section{Models\label{sec:sup:Models}}

\subsection{Topological surface state dispersion\label{sec:sup:ARPES}}
Peaks associated with the topological surface state in the momentum distribution curves as measured by ARPES\cite{ZhuPRL2011,ZhuPrivate}  were fit  along the $\Gamma$-K and $\Gamma$-M directions at many binding energies ranging from the valence band to well above the conduction band edge. Fits to the dispersion where energy is measured from the Dirac point give $E|_{\Gamma-K}=1.7 k+13.5 k^2-4,600 k^6$ and $E|_{\Gamma-M} =2.2 k+8.2 k^2+4,200 k^6$ where $E_F$ is in eV and $k$ is in ${\AA}^{-1}$. There is little difference between the two dispersions from hexagonal warping\cite{Fu_Warping2009} since our range of Fermi level is from the Dirac point to just above the conduction band edge. The average of the two dispersions is used in the analysis and is given by $E_{avg}=1.9 k+12.6 k^2+2,300 k^6$.  The TSS Fermi velocity $v_F= dE/d(\hbar k)$, cyclotron mass $m_c= \hbar k_F /v_F$, and carrier density $n=k_F^2/(4 \pi)$ are used in analyses.

\subsection{Thomas-Fermi screening Model\label{sec:sup:T-Fmodel}}

\begin{figure}[h]
\includegraphics[scale=.4]{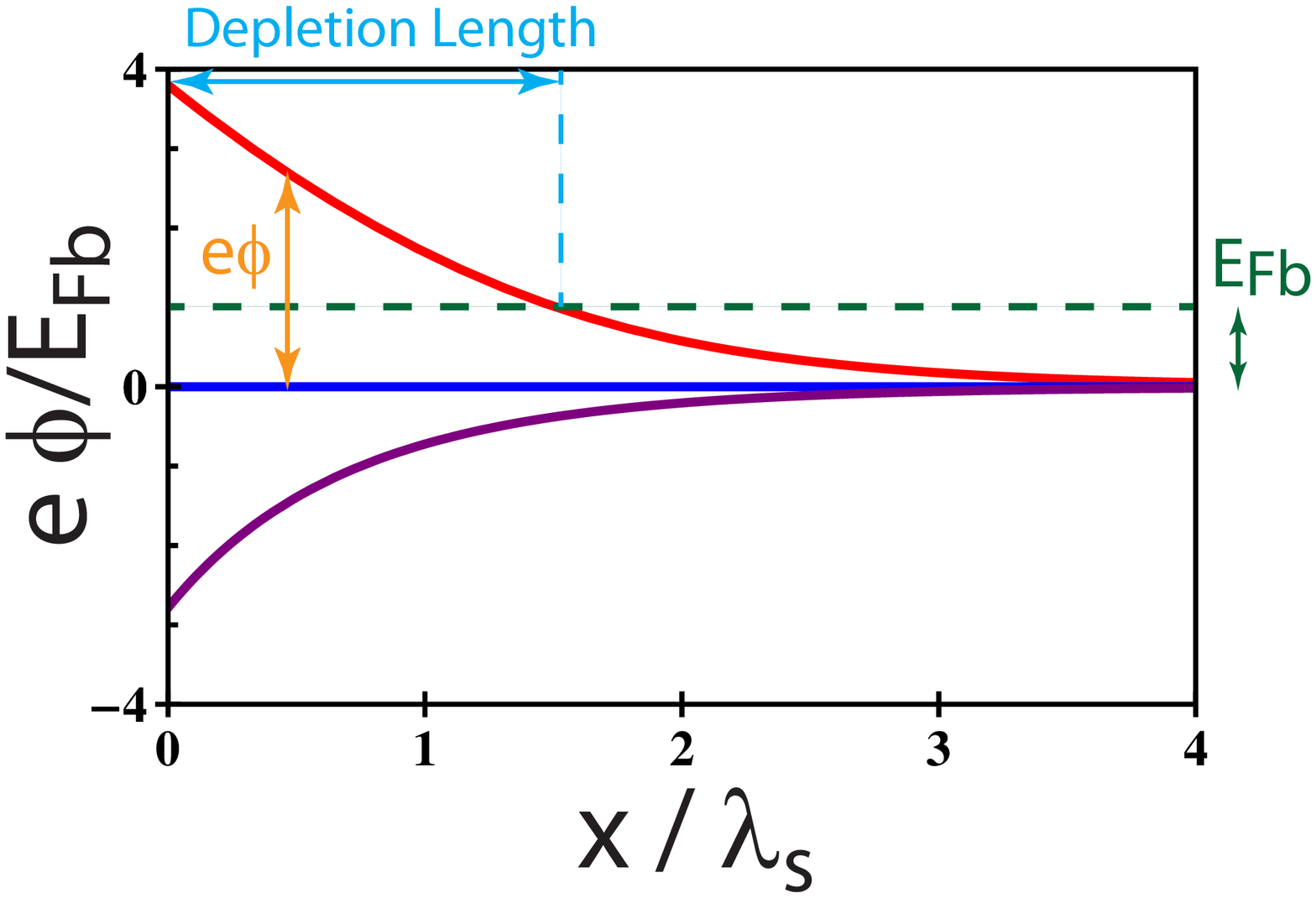}
\caption{\label{fig:Supbandbending} Calculated band bending effects of the conduction band edge (red, blue, purple) in a semi-infinite slab of topological insulator. $E_{Fb}$ (dashed green) is the Fermi level deep in the bulk, $\phi$ is the potential due to an externally applied electric field, $\lambda_s$ is the screening length, and $x$ is the distance away from the gated surface. Depletion onset occurs when $e \phi=E_{Fb}$. For larger $\phi$ (red), a depletion layer forms separating the topological surface state from the bulk carriers. Zero $\phi$ (blue) defines flat band where the background ion density equals the electron density. For negative $\phi$ (purple), the conduction band edge bends downward creating an accumulation layer.}
\end{figure}

Knowing the capacitance, a gate transfers a determined amount of net charge to or from a topological insulator. The net charge is distributed between the topological surface state and the bulk. Deriving this proportion requires solving the general band bending problem in a topological insulator due to screening of potentials.\cite{Stanescu2012}

In the Thomas-Fermi screening model, the potential satisfies Poisson's equation which we solve for a semi-infinite slab of topological insulator subject to an externally applied electric field. Figure \ref{fig:Supbandbending} shows band bending of the conduction band edge as a function of depth with a Fermi level $E_{Fb}$ defined away from the screened region.

\begin{equation*}\label{PoissonEq}
  \frac{\partial^2 \phi(x)}{\partial x^2}=\frac{\rho(x)}{\epsilon}
\end{equation*}

The net electrostatic charge density is related to the net ionic background density $n_0$ and the carrier density of electrons n(x) by $\rho (x) = e (n_0 -n(x))$
where $n_0 = \frac{1}{3 \pi^2} (2 m E_{Fb}/\hbar^2)^{3/2}$ and $n(x) = \frac{1}{3 \pi^2} (\frac{2 m}{\hbar^2} (E_{Fb} - e \phi(x)))^{3/2}$

The differential equation can be written in the following general form:
\begin{equation*}\label{Sdiffequ}
  \frac{\partial^2 S}{\partial y^2}=(1-(1-S)^{3/2})
\end{equation*}
where $S = \frac{e \phi}{E_{Fb}}$, $y=\frac{x}{\lambda_s}$, and $\lambda_s^2 = \frac{\epsilon E_{Fb}}{e^2 n_0}$. $\lambda_s$ defines the screening length and $\epsilon \approx 100$ for Bi$_2$Se$_3$.\cite{Sushkov_PRB2010} Defining $  E_S \equiv \frac{d S}{d y}$ and integrating, the second order differential equation is converted into a first order one. For the specific case of a semi-infinite slab where $E_S$ and $S$ are zero deep inside the topological insulator, the solution is simply given by:
\begin{equation*}
  E_S = \pm (2 S + \frac{4}{5}(1-S)^{5/2} - \frac{4}{5})^{1/2}
\end{equation*}
Where the + (-) sign is for the case of band bending down (up), or equivalently $S(0)<0$ ($S(0)>0$). Depletion onset occurs when $S(0)=1$. Either $E_S$ or $S$ on the inside surface at $y=0$ is determined given an externally applied gate voltage and the uncompensated charge associated with the topological surface state whose dispersion is characterized by ARPES.

\subsection{Optical Models\label{sec:sup:opticalmodel}}

The general optical models which were used in simulating the various data sets are described. Approximate formulas relating to various optical measurements are presented first which emphasize the physical concepts.

The normal-incident transmission $T$ for a thin film with multiple contributions to the conductivity $\sigma$ relative to the transmission of the substrate $T_0$ is given by $\frac{T}{T_0}=|\frac{n_{sub} + 1}{n_{sub} + 1 + \sum {y_i} }|^2$ where $ y_i = Z_0 \sigma_i$ is the surface admittance, $n_{sub}$ is the index of refraction of the substrate, and $Z_0$ is the impedance of free space.

Fourier transform spectroscopy (FTIR) measurements performed in zero magnetic field are described within a Drude model defined by $ \sigma_{xx} = \frac{\omega_{ps}/Z_0}{\gamma - \imath \omega}$ where $\omega_{ps} = Z_0 n e^2/m$. $\sigma_{xx}$ is the longitudinal conductivity, $\omega_{ps}$ is the two dimensional plasma frequency, $\gamma$ is the scattering rate, $\omega$ is the radiation frequency, $n$ is the two dimensional carrier density, $e$ is the electronic charge, and $m$ is the effective mass.

Cyclotron resonance (CR) transmission measurements are performed at fixed frequency with normally incident circularly polarized light as a function of applied magnetic field. The surface admittance terms are most simply given in the circular polarization basis where $ \sigma^{\pm}  = \frac{\omega_{ps}/Z_0}{\gamma - \imath (\omega \pm \omega_c)}$ where $\omega_c = e B / c \, m_c$ is the cyclotron frequency, $m_c$ is the cyclotron mass, and $\sigma^{\pm}$ is the conductivity in the circular polarization basis. In the limiting case $y_i << n_{sub} +1$, the differential transmission reduces to $ \Delta T^\pm = -\frac{2 T_0}{n_{sub} + 1} Re(\sum {Z_0 \, \Delta \sigma^\pm_i})$. Differential cyclotron resonance measurements ($\Delta$-CR) can be thought of as a sum of differences of Lorentzians. The polarity and value of the B-field resonance  determines the sign and mass of the carriers.

The complex Faraday angle is defined as   $\tan{\theta_F} = \imath \frac{t^+ - t^- }{t^+ + t^-} = \imath \frac{\sum{Z \sigma^i_{xy} d}}{1+\sum{Z \sigma^i_{xx} d}}$ where $\tan{\theta_F}\approx \theta_F$, $t^{\pm}$ are the transmission Fresnel coefficients in the circular polarization basis, and $2 \imath \sigma_{xy}=\sigma^+ - \sigma^-$.\cite{Jenkins_RSI_2010} Im($\theta_F$) is related to the circular dichroism similar to cyclotron resonance measurements. The real part, thought of as the angle of polarization rotation, is related to the reactive part of the conductivity. When the scattering rate is much less than the cyclotron frequency, $Im(\theta_F)$ ($Re(\theta_F)$)  shows a simple Lorentzian resonance (antiresonance)  centered at $B_{res}$. Contrary to CR measurements, n and p-type carrier resonances will simultaneously appear in both polarities of magnetic field but with opposite sign.

The actual model used to simulate the data is derived from a general solution to the transmission Fresnel coefficient for light normally incident on a stack of slabs which may have multiple conducting films at interfaces. The formalism is given in Reference \citenum{Jenkins_PRB2010}, Appendix I.

Since the parylene coating and In$_{2}$Se$_{3}$ capping layer are thin (compared to wavelength) non-absorbing dielectric films in our frequency range, they do not effect the optical signals so we omit them. The Bi$_2$Se$_3$ sample is a conducting film with optical absorptions due to phonons. Therefore the first boundary is between air and the Bi$_2$Se$_3$ slab, and between them is a NiCr film (used as a gate and characterized by the dc sheet resistance $R_g$). All contributions from all the free carriers in the  Bi$_2$Se$_3$ film are lumped into $\sigma^{\pm}$. The second interface is between the Bi$_2$Se$_3$ and sapphire substrate.  The last interface is sapphire into air with a NiCr film, an anti-reflection coating characterized by sheet resistance R$_{AR}$.

The Fresnel transmission coefficient $t^{\pm}$ are derived for the entire stack of slabs.  All optical quantities presented in this paper are calculated from this expression.

The cyclotron resonance transmission is given by $T^+[V_g] = t^+ \overline{t^+}$. The FTIR transmission spectra, although unpolarized, is given by the same expression with $B=0$ and $\sigma^+$ replaced by the longitudinal polarization expression $\sigma_{xx}$.

For the differentially gated measurements, $\Delta$-CR is given by
\begin{equation*}\label{eqDifCycResSig}
\frac{\Delta T^+}{T_{avg}} = \frac{T^+[V_g+\Delta V/2] - T^+[V_g-\Delta V/2] }{\frac{1}{2}(T^+[V_g+\Delta V/2] + T^+[V_g-\Delta V/2] )}
\end{equation*}
Unpolarized gated FTIR measurements are reported normalized to the zero gate transmittance and is given by (where B is set to zero)$\frac{T^+[V_g]}{T^+[V_g=0]}$. $\Delta \theta_F$ is given by $\Delta \theta_F[V_g] = \theta_F[V_g+\Delta V/2] - \theta_F[V_g-\Delta V/2]$.

There are two prominent phonons in Bi$_2$Se$_3$ over our spectral range of interest which can be written in terms of the complex index of refraction $n_{BiSe}=\sqrt{\epsilon}$ where $\epsilon=\epsilon_0+\sum\limits_{i=1}^2 \epsilon_i$  and $\epsilon_i=\frac{\Omega_i^2}{\omega_i^2-\omega^2-i \omega \Gamma_i}$. The phonon parameters for epitaxial films are experimentally determined to be $\epsilon_0 = 25.6$, $\Omega_1 = 615 \textrm{ cm$^{-1}$}$, $\omega_1 = 63.1 \textrm{ cm$^{-1}$}$, $\Gamma_1 = 1.8 \textrm{ cm$^{-1}$}$, $\Omega_2 = 80 \textrm{ cm$^{-1}$}$, $\omega_2 = 133 \textrm{ cm$^{-1}$}$, and $\Gamma_2 = 2 \textrm{ cm$^{-1}$}$. Similar values for bulk crystals were previously reported.\cite{Sushkov_PRB2010}

The free carrier response of the Bi$_2$Se$_3$ film is modeled as the sum of three Drude terms $ \sigma^{\pm} =  \sum\limits_{i=1}^3 \frac{\omega_{ps,i}/Z_0}{\gamma_i - \imath (\omega_i \pm \omega_{c,i})}$ where $\omega_{c,i} = e B / c m_{c,i}$ is the cyclotron frequency, $\omega_{ps,i}= Z_0 n_i e^2 / m_{c,i}$ is the two dimensional plasma frequency, $\gamma_i$ is the inverse transport lifetime, $m_{c,i}$ is the cyclotron mass, $n_i$ is the two dimensional carrier density, and Z$_0$ is the impedance of free space.

NiCr films are used as a gate and antireflection (AR) coating since NiCr has a very high scattering rate. Therefore, the admittance is frequency independent at THz frequencies. The optical response is then completely characterized by the dc sheet resistance where $\textrm{R$_{AR}$} = 275 \textrm{ $\Omega$}$, and $\textrm{R$_g$} = 400 \textrm{ $\Omega$}$. Other parameters are the index of refraction of sapphire $\textrm{n$_{Saph}$}=3.1$ and vacuum $\textrm{n$_0$} = 1$, the thickness of the sapphire substrate $\textrm{d$_{Saph}$}=.05 \textrm{ cm}$, and the Bi$_{2}$Se$_3$ film thickness $\textrm{d$_{BiSe}$} = 60 \textrm{ nm}$.

\section{Discussion\label{sec:sup:Discussion}}

\subsection{Intuitive description of  Figure \ref{fig2}(d), $\Delta$-CR model\label{sec:sup:intuitivedesc}}

\begin{figure}
\includegraphics[scale=.35]{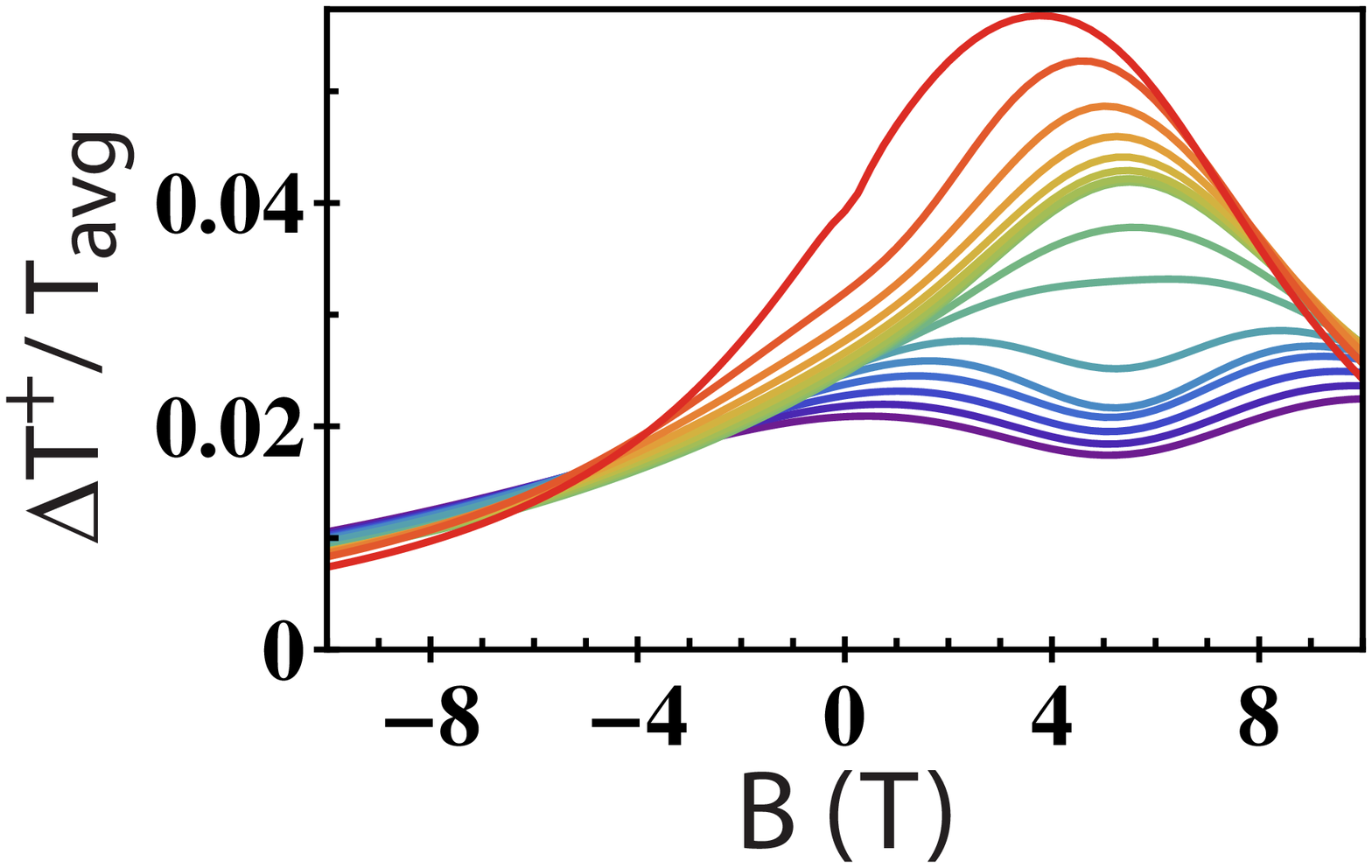}
\caption{\label{fig:SupDCRModelDemo} \textbf{$\Delta$-CR model with constant $\gamma_{TSS}$} The $\Delta$-CR model is the same as in Figure \ref{fig2}(d) with exactly the same parameters, but the top topological surface state scattering rate is set to a constant 9.4 THz, independent of gate voltage.}
\end{figure}

The modeled optical signals in relation to the Drude parameters can be qualitatively understood most easily by considering the $\Delta$-CR model curves of Figure 2(d). A cyclotron resonance is a Lorentzian-like absorption where $n/m_c$ defines the weight,  $\gamma$ defines the width, and $m_c$  defines the center of the resonance via $B_{res} = (c\,\omega / e) m_{c} $. When a gate voltage is applied, these parameters can change.  A single $\Delta$-CR curve is  the difference between pairs of Lorentzians, a pair for the TSS and MTB contributions. Since the constant BSAL conductivity component is not modulated by the gate, it may be left out of the following qualitative discussion of the $\Delta$-CR model.

Near zero gate voltage, the mass and scattering rate of the TSS and MTB carriers do not change much with gate voltage as shown in Figure \ref{fig3}. Only the weights of the resonances change. Since $\gamma_{ TSS}>>\gamma_{MTB}$ and the change in carrier density $\Delta n_{ TSS} \sim \Delta n_{MTB}$, the resulting line shape appears Lorentzian centered on the mass of the MTB contribution. As the gate is increased and the surface Fermi level moves higher, the dominant response is the MTB contribution that has a growing scattering rate and carrier density. Taking the difference between two Lorentzians that have slightly different widths ($\gamma$) will cause a suppression of the peak at the resonant B-field. A slight simultaneous increase in mass shifts one of the resonances in B-field relative to the other causing a slight skew in the difference curve. In this case, the part of the differential cyclotron resonance occurring at low B-fields are slightly more suppressed than the higher fields due to the changing MTB mass. The gradual increase of the large TSS scattering rate causes a  suppression of the resonant peak over a broad range of B-field.

As one gates from zero to negative gate voltages toward the Dirac point, the only MTB Drude parameter that changes is the carrier density. Therefore, the $\Delta$-CR of the MTB carriers maintain an approximately constant Lorentzian$-$like line shape.

The TSS response is superimposed on the MTB background. The mass and number density decrease with surface Fermi level consistent with the ARPES measured Dirac dispersion. The model results for a constant TSS  scattering rate are shown in Figure \ref{fig:SupDCRModelDemo} where all other model parameters are exactly the same as reported in Figure \ref{fig3}. In this case, the TSS contributions to $\Delta$-CR are a series of skewed Lorentzian-like peaks whose weight increases with decreasing surface Fermi level, and whose resonant location ($B_{res}\sim m_c$) decrease towards zero B-field as expected from a Dirac cone dispersion. The increase of differential spectral weight can be seen for the case of an ideal Dirac cone where $SW = n e^2/m_c \sim n^{1/2}$, so $d(SW)/dn$ increases as the Dirac point is approached. The peak-dip-hump structure that is present in the data in Figure \ref{fig2}(a) is conspicuously missing in Figure \ref{fig:SupDCRModelDemo}.

The very rapid change of the TSS scattering rate ocurring at approximately -70 V gives rise to the peak-dip-hump structure. The depth of these modulations in the $\Delta$-CR curves of Figure \ref{fig2}(d) are extremely sensitive to the rate of change of the scattering rate, and therefore extremely sensitive to the shape of the scatterring rate curve shown in Figure \ref{fig3}(d).

Since the model does not include potential fluctuations, the modeled $\Delta$-CR peaks for the TSS contribution at the highest negative gate voltages can be thought of as just the CR peak of the higher surface Fermi level since very little relative spectral weight exists in the lower surface Fermi level peak. This is most vividly illustrated by considering the hypothetical scenario where the lowest surface Fermi level is actually at the Dirac point where there are no carriers. The difference curve then would be exactly the TSS CR curve at the higher surface Fermi level, superimposed on the MTB contribution.

An understanding of all the $\Delta$-CR curves can be summarized. The MTB contribution is the dominant response at positive gate voltages. At negative voltages, the MTB contribution to $\Delta$-CR is roughly constant since only $n$ is changing. The peak-dip-hump structure that develops is associated with the TSS contribution primarily due to the rapidly changing scattering rate. The shifting of this peak-dip-hump structure with B-field is a result of the rapidly changing mass expected from a Dirac cone. At the highest negative gate voltages, a single TSS cyclotron resonance is superimposed on a constant $\Delta$-CR MTB Lorentzian-like peak.

\subsection{Carrier distributions, and energy estimates of potential fluctuations and the conduction band edge \label{sec:sup:BSALexistence}}
dc transport measurements on uncapped films, grown in exactly the same manner as our film, show a 2-D carrier density $\thickapprox 4 \times 10^{13}$ e/cm$^{2}$ over a wide range of film thicknesses, 10 to 250 nm. The thickness independence of the carrier density indicates the bulk contribution is small compared to contributions from the two accumulated surfaces (topological surface states and bulk accumulation layers). The bulk carrier density is measured to be $\lesssim 5 \times 10^{17}$ cm$^{-3}$.\cite{Oh_arxiv2011} The cyclotron mass is $\sim0.2$ m$_0$ (taken directly from the THz Faraday and cyclotron resonance data). The screening length $\lambda_s$ depends weakly on bulk density, so it is $\sim 11$ nm. The bulk density corresponds to a bulk Fermi level of $\sim11$ meV.

For comparison, this carrier density distributed uniformly over a 60 nm film results in a 2-D carrier density of only $\lesssim3 \times 10^{12}$ cm$^{-2}$. Single-fluid fits to the zero-gate data in Figure \ref{fig1} (blue curves) give the density in the film, where the FTIR data gives $1.2\times10^{13}$ cm$^{-2}$ and $\theta_F$ data gives $1.1\times10^{13}$ cm$^{-2}$. Using the entire analysis with a three-fluid fit to the zero-gate data that incorporates results from the differential measurements, the total carrier density at zero gate is $1.44 \times 10^{13}$ cm$^{-2}$ distributed between the BSAL contribution of $1.0 \times 10^{13}$ cm$^{-2}$ and the top surface contribution of $4.4 \times 10^{12}$ cm$^{-2}$.

The gated optical measurements, independent of the dc characterization, show that most of the charge measured at zero gate resides in the bottom section of the film. The very small measured mass at large negative gate voltage in the $\Delta$-CR data indicates the surface state has a Fermi level very close to the Dirac point at -170 V. At this voltage, a depletion region necessarily exists at the top surface, as depicted in Figure \ref{fig:Supbandbending}, with depletion charge $e n_d$. The depletion charge is calculated by solving the Thomas-Fermi screening model, giving $n_d = n_0(\lambda_s^2 + \frac{2 \epsilon}{n_0 e^2} E_c)^{1/2} - n_0 \lambda_s$ where $E_c$ is the energy of the conduction band edge above the Dirac point and $n_0=5\times 10^{17}$ cm$^{-3}$ is the bulk density consistent with dc characterizations.\cite{Oh_arxiv2011}

From the Dirac point to $V_g=0$,  $4.4\times10^{12}$ cm$^{-2}$ carriers have been added to the film filling this depletion layer and transferring carriers into the TSS and bulk.  Therefore, the net carrier density at the top surface is $n_s= 4.4\times10^{12} \text{cm}^{-2} - n_d < 4.4\times10^{12} \text{cm}^{-2}$.  The depletion charge for a band bending potential of 190 meV is $2.7\times10^{12}$ cm$^{-2}$, so there are $1.7\times10^{12}$ cm$^{-2}$ excess carriers on the top surface.  The remaining  $>10^{13}$ cm$^{-2}$ carriers measured at zero gate are in the bottom surface and bulk.

For the shifted Dirac cone deduced from the optical data, which is one of the main findings of the paper, the surface carrier density is somewhat larger at zero gate. However, it is necessary to estimate the surface Fermi level of the conduction band edge located at $-70$ V (while the Dirac point remains at -170 V) before it is possible to estimate the net charge $e n_s$ on the top surface at zero gate.

In the case of this shifted Dirac cone, to gate from the Dirac point to the conduction band edge requires $n_g=2.6 \times 10^{12}$ cm$^{-2}$ carriers to fill the surface state up to the conduction band edge with $n_{fss}$ carriers as well as fill the depletion layer with $n_d$ carriers. $n_{fss}$ is found from the ARPES dispersion and the Fermi level of the conduction band edge, $E_c$. Solving $n_g=n_d(E_c)+n_{fss}(E_c)$ gives $E_c=80$ meV where $n_{fss} = 1\times 10^{12}$ cm$^{-2}$ and $n_d=1.6\times10^{12}$ cm$^{-2}$.

By solving the full Thomas-Fermi screening model with the conduction band edge set at 80 meV above the Dirac point, the surface Fermi level $E_F$ as a function of gate voltage can be calculated, similar to Figure \ref{fig3}(b). Such a treatment provides estimates of the potential fluctuation energies. As specified in the main text, the onset of potential fluctuations near the Dirac point occurs at -130 V. This corresponds to 50 meV. Taking a derivative of the scattering rate curve shown in Figure \ref{fig3}(d) shows a gaussian-like structure centered on the step at -70 V where the FWHM points are at -100 V and -40 V, corresponding to an energy width of 50 meV. The 1/2-width is compared with the Dirac point potential fluctuation onset energy in the main text.

At any rate, the net surface charge for the case of the shifted Dirac cone when gating from the Dirac point to $V_g=0$ is $n_s= 4.4\times10^{12} \text{ cm}^{-2} - n_d < 4.4\times10^{12} \text{ cm}^{-2}$.  The excess charge on the surface is $n_s=2.8 \times 10^{12}$ cm$^{-2}$.  Again we can conclude that most of the carriers reside on the bottom surface of the film.

Using the dc characterized number density is not necessary to conclude that most carriers in the film reside on the bottom. Larger flat band bulk densities $n_0$ lead to larger $n_d$, and therefore less carriers on the top surface at zero gate.

For a bulk density of $5\times 10^{17}$ cm$^{-3}$, whether the conduction band edge is at 80 or 190 meV, the depletion region extends across at least half of the film at $V_g=-170$ V. For larger bulk densities, the depletion length becomes smaller. The BSAL Drude conductivity contribution, the component that is not modulated by the top gate by definition, can involve the back surface accumulation layer and any bulk contribution not depleted at $V_g=-170$ V.

It is not \textit{a priori} obvious that three Drude terms should suffice in the model since numerous other conductivity  contributions could be associated with bulk and accumulated carriers. The modulated top bulk (MTB) region described as one Drude term is conceptually appealing. Gating from the Dirac point to the onset of accumulation only modulates the number density of the bulk carriers associated with flat-band of the film, providing an explanation as to why the MTB scattering rate and mass are gate independent. As the carriers begin to accumulate at positive gate voltages, some of the bulk states gradually change into confined states at the top surface increasing the bulk/surface scattering channels while also increasing bulk scattering from surface defects. The mass of the accumulated carriers also gradually increases, heading toward the bigger mass found on the larger accumulated bottom surface.

For the bottom surface accumulation layer (BSAL), there may exist many subband states associated with the bulk accumulation, and the lowest subbands are likely Rashba split.\cite{Bianchi_Hofmann_2010,ZhuPRL2011} However, multiple bottom surface carrier contributions are not discernable in the zero-gate measurements nor the differential optical measurements that are not very sensitive to the details of the BSAL.

The three-term Drude model impressively reproduces the features of all the data sets with only three free parameters allowing clean extraction of the TSS parameters.

\subsection{Cyclotron and SdH mass\label{sec:sup:IBBSALMasses}}

The measured cyclotron mass of the bulk MTB carriers ($0.21$) near the conduction band edge as well as the BSAL mass ($0.25$) may at first appear large. The Shubnikov-de Haas (SdH) measured mass in bulk crystals is $0.13-0.15$ near the conduction band edge with a carrier density of $\sim 10^{17}$ cm$^{-3}$, and shows no discernible non-parabolicity up to $\sim 5 \times 10^{19}$ cm$^{-3}$. \cite{Köhler1973, AnalytisPRB2010, EtoAndo2010, Kulbachinskii1999, Hyde1973, Navrátil2004} This behavior of the SdH mass may seem peculiar considering that, as is generally known, the band dispersion in small-gap semiconductors is ordinarily non-parabolic causing the band mass, and therefore the cyclotron mass, to increase with surface Fermi energy.

For systems with degenerate carriers, however, many-body interactions can complicate direct comparison of measured masses.\cite{PlatzmanWolff1973} In SdH (and ARPES) the measured cyclotron frequency differs from the band value due to self-energy corrections coming from the electron-electron interaction
due to the Coulomb interaction and the Frohlich interaction due to the polar phonon.

\begin{figure*}
\includegraphics[scale=.85]{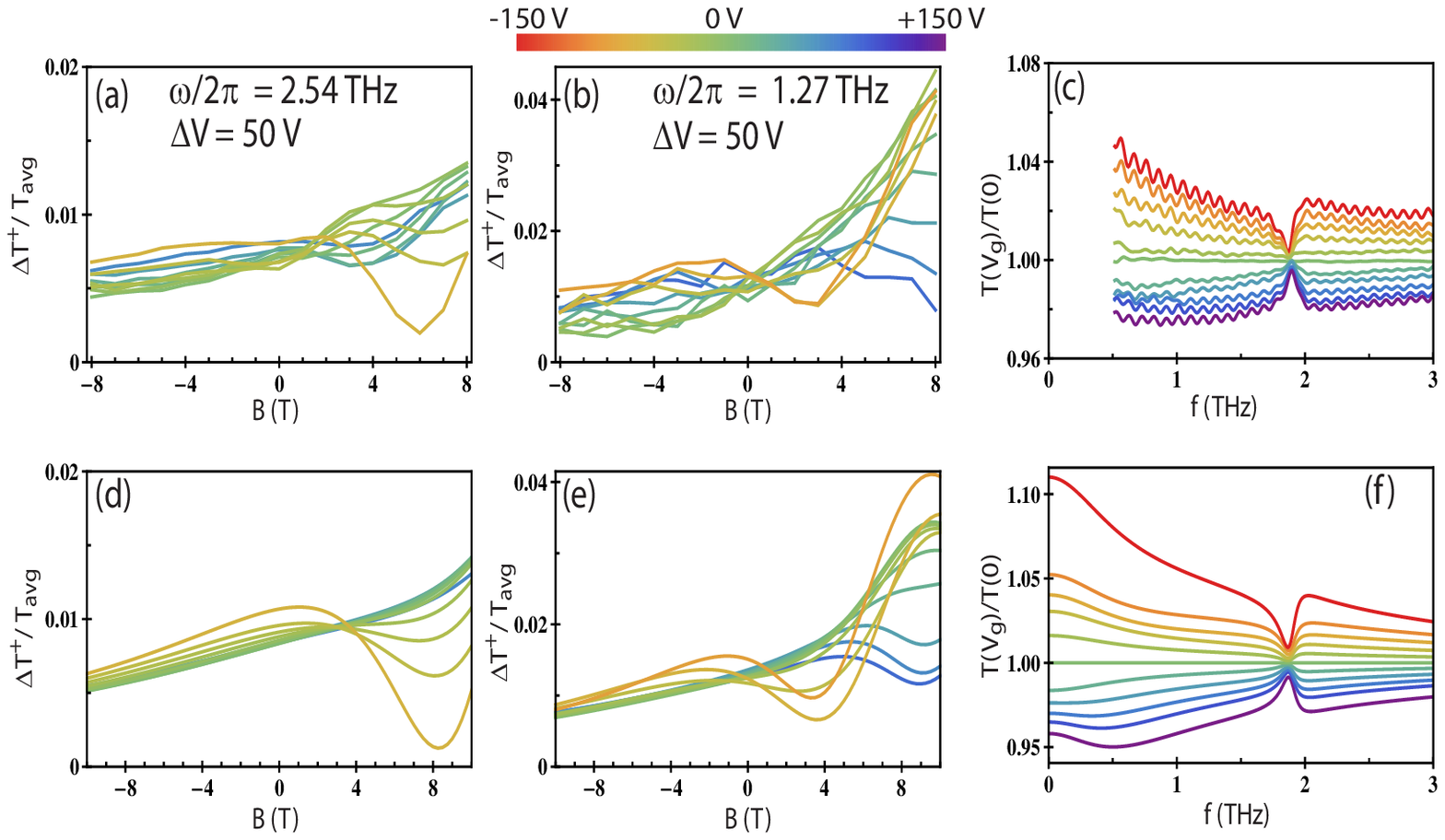}
\caption{\label{fig:SupDVgDataf} \textbf{Gated optical transmission measurements addenda} (a-b) 10K $\Delta$-CR measurements are the same as described in the caption of Figure \ref{fig2} except measured at different frequencies labeled on each graph (c) 6 K zero magnetic field FTIR transmission measured at gate voltage $V_g$ normalized to the zero gate value. (d-f) the corresponding modeled optical responses using the same parameters reported in Figure \ref{fig3} that reproduce the data of Figures \ref{fig2}(a-c) and Figures \ref{fig1}(c,d).}
\end{figure*}

This effect is dependent on carrier density and may differ for 2D and 3D electron systems. In the case of cyclotron resonance, final state interactions associated with the photo-generated electron-hole pair (exciton effect) can lead to a shift of the cyclotron resonance frequency compared with the SdH (or deduced from ARPES).  This excitonic shift tends to compensate for the self-energy effects resulting in a cyclotron resonance that tends to be closer to the bare-band value.  The Kohn thereom\cite{McKnight1980, Verdun1974, VerdúnDrew1976} shows that this compensation is exact for pure Coulomb interactions in an isotropic 2D electron liquid or a 3D electron liquid at q=0.

Unfortunately, there has not been much work comparing SdH and magneto-optical data to sort out these many-body effects in Bi$_2$Se$_3$ or related materials.  In our earlier work, \cite{Jenkins_PRB2010, Sushkov_PRB2010} the bulk magneto-transmission yielded a band edge mass of $0.15$. However, in this measurement which is dominated by magneto-plasma effects, $q\neq0$ so that the cyclotron mass should be compared with the SdH mass with which it agrees.

Measurements on PbTe and bismuth metal illustrate these many-body interaction effects.\cite{McKnight1980, Verdun1974, VerdúnDrew1976}  We note that $r_s$, the ratio of the electronic potential to kinetic energy and a general measure of the strength of interactions, is larger for Bi$_2$Se$_3$ ($r_s=0.2$) compared with bismuth ($r_s=0.1$) so that larger exciton shifts are expected.

In view of these considerations the observation of a $\sim 0.2$ mass for the MTB channel is not surprising.  In these thin films, $q=0$ and so final state interactions effects are relevant. In addition, the mass observed for the zero-gate optical response of Figure \ref{fig1}, which is dominated by the BSAL, appears to be even larger in our work as well as the work of Armitage \emph{et al}.\cite{AguilarPRL2012}

These cyclotron mass issues deserve further study in these topologically interesting materials.

\subsection{topological interface states\label{sec:sup:DPshift}}

The interface between two insulating materials with band gaps inverted with respect to each other can support topologically protected interface states.\cite{KorenmanDrew1987,QiZhangRMP2011}  For the ARPES measurements on bulk topological insulating materials the vacuum takes the part of the topologically trivial insulator.   The more general case has been solved within the two band model using $k \cdot p$ perturbation theory in reference \citenum{KorenmanDrew1987}.  This theory shows that the Dirac spectrum is modified as a function of the gap of the trivial insulator and any potential step at the interface.  Thus for a trivial insulator with a gap of 1.2 eV in contact with a topological insulator with a bulk gap of 300 meV in the presence of a potential step of 0.3 eV the Dirac point of the topological surface state shifts by 60 meV.  This potential step can result from differences in the work function of the two materials or the presence of a dipole layer at the interface.  While not enough is known about the In$_2$Se$_3$/Bi$_2$Se$_3$  interface to predict the potential step,\cite{Drapak_InSeW2002,Sakalauskas_InSeW1998,BerntsenARPESBi2Se3W2012} this calculation demonstrates that a shift of the right magnitude is expected to occur with reasonable values of a potential step.  A more accurate theory together with experimental determination of the potential step is needed to confirm this result.

\section{Gated optical data addenda\label{sec:sup:moredata}}

Figures \ref{fig:SupDVgDataf}(a,b) report $\Delta$-CR data that are similar to Figure \ref{fig2}(a) except the measurements were performed at different frequencies. Figures \ref{fig:SupDVgDataf}(d,e) are the modeled optical responses that use the same parameters as reported in Figure \ref{fig3}.

Differential Fourier transform spectroscopy data ($\Delta$-FTIR) and modeled response are reported in Figures \ref{fig:SupDVgDataf}(c) and (f). The oscillations present in the data are from Fabry-Perot interference occurring in the substrate due to an imperfect antireflection coating. The increase in transmission with negative gate voltage is evidence of dominant n-type carriers.  If a single Drude n-type carrier existed and a gate changed only the carrier density, then the data set would be fully symmetric about zero gate bias. The asymmetry is an indication of a gate-dependent scattering rate and/or mass.

In the case of the model in Figures \ref{fig:SupDVgDataf}(f), the main effect is produced by the gate-dependent MTB carrier scattering rate with some effects from the TSS carriers. For increasing positive gate voltages, the MTB and TSS carriers have progressively larger masses and scattering rates giving rise to progressively smaller changes in conductivity.  For negative gate voltages, the opposite is true so the changes in conductivity are larger.

A 7 mm spot size was used for the FTIR measurements. Such a large spot size relative to the sample allows the maximum of throughput power required to measure to low frequencies. However, some leakage light occurred through ungated parts of the film due to the shadow masking procedure. The measured $\Delta$-FTIR signals are therefore reduced. Although the photometrics differ from the model, the qualitative behavior is reproduced. The apparent asymmetric response observed in the transmission associated with the phonon resonance in Figures \ref{fig1}(c) and \ref{fig:SupDVgDataf}(c) is reproduced by a simple Lorentzian dielectric response function and follows from the divergence of the real part of $\epsilon$ associated with the transverse-optical phonon at $f\approx 1.9$ THz.


\end{document}